\documentclass[pra,twocolumn,superscriptaddress,10pt,noshowpacs]{revtex4-1}
\usepackage[english]{babel}
\usepackage[T1]{fontenc}
\usepackage[utf8]{inputenc}
\usepackage{graphicx,epstopdf}
\usepackage{amsmath,amssymb}

\usepackage{amsfonts}
\usepackage{color}
\usepackage{latexsym}
\usepackage{caption}
\usepackage{subcaption}
\usepackage{times,txfonts}

\newcommand{\meio}{{}^1\!/{}_{\!2}}

\newcommand{\beq}{\begin{equation}}
\newcommand{\eeq}{\end{equation}}
\newcommand{\bea}{\begin{eqnarray}}
\newcommand{\eea}{\end{eqnarray}}
\begin{document}

\title{On quantum traversability of wormholes}

\author{J. Furtado\footnote{E-mail:job.furtado@ufca.edu.br}}\affiliation{Universidade Federal do Cariri, Centro de Ci\^encias e Tecnologia, 63048-080, Juazeiro do Norte, CE, Brasil.}

\author{C. R. Muniz\footnote{E-mail:celio.muniz@uece.br}}\affiliation{Universidade Estadual do Cear\'a, Faculdade de Educa\c c\~ao, Ci\^encias e Letras de Iguatu, 63500-000, Iguatu, CE, Brazil.}

\author{M. S. Cunha\footnote{E-mail:marcony.cunha@uece.br}}\affiliation{Universidade Estadual do Cear\'a, Centro de Ci\^encias e Tecnologia, 60714-903, Fortaleza, CE, Brazil.}

\author{J. E. G. Silva\footnote{E-mail:euclides.silva@ufca.edu.br}}\affiliation{Universidade Federal do Cariri, Centro de Ci\^encias e Tecnologia, 63048-080, Juazeiro do Norte, CE, Brasil.}


\begin{abstract}
In this paper we study the possibility of non-relativistic quantum particles to traverse the generalized Ellis-Bronnikov wormholes by considering quantum effects, such as tunneling. We have used the generalized Ellis-Bronnikov wormhole metric and found that for $n=2$ we have a single barrier shaped effective potential centered at the throat of the wormhole for any value of orbital angular momentum. For $n\neq2$ we have a symmetric double barrier shaped potential when the orbital angular momentum is zero and a single barrier for nonzero angular orbital momentum. Analytical solutions for the Schr\"{o}dinger equation in the generalized Ellis-Bronnikov spacetime could be found only for $n=2$. Such solutions were given in terms of the confluent Heun functions. Finally, by using a delta-barrier approximation we could find the transmission and reflection coefficients for a non-relativistic particle to traverse the generalized Ellis-Bronnikov wormhole.
\end{abstract}

\maketitle

\section{Introduction}

The idea of a wormhole, i.e., a bridge connecting two asymptotically flat regions of the same universe or two different universes was first hypothesized in \cite{Einstein:1935tc} and it was known as the Einstein-Rosen bridge. However, wormholes traversability in a more general way was only studied more than fifty years later by Morris and Thorne \cite{Morris:1988cz}. An important feature of wormholes in Einstein's theory of gravity is that its traversability requires exotic matter \cite{Morris:1988cz}, being possible to have wormholes with phantom as energy source \cite{Sushkov:2005kj, Lobo:2005us} or even with Casimir energy \cite{Garattini:2019ivd, Jusufi:2020rpw, Alencar:2021ejd, Oliveira:2021ypz, Carvalho:2021ajy}. Therefore the search for traversable wormholes in modified theories of gravity \cite{Richarte:2007zz, Matulich:2011ct, Richarte:2009zz, MontelongoGarcia:2011ag, Ovgun:2018xys, Lessa, Chew:2016epf, Chew:2018vjp} without the requirement of exotic matter became an intense topic of research in the literature.

The first traversable wormhole solution was found by Ellis and Bronnikov \cite{Ellis:1973yv, Bronnikov:1973fh}, a few years before the seminal work of Morris and Thorne. In his work, Bronnikov realized, with evidence, that the Ellis drainhole is geodesically complete, without event horizons, with free singularity and with traversability independent of direction \cite{Ellis:1973yv, Bronnikov:1973fh}. On top of that, knowing that the wormhole's scalar field source is phantom-like, of course, all energy conditions of General Relativity (GR) are violated.

Some time ago a class of wormhole solutions based in the Ellis-Bronnikov spacetime \cite{Ellis:1973yv, Bronnikov:1973fh} was proposed in \cite{Kar:1995jz}, as an attempt to get around the problem of exotic matter. These solutions were called generalized Ellis-Bronnikov wormholes, which were studied in the context of general relativity in \cite{Kar:1995jz}. The authors investigated the resonances of the propagation of scalar waves in this family of wormholes. Further, they showed that these classes of wormholes can only be supported by exotic matter, due to the violations of the null energy condition and the weak energy condition in the classical context. More recently, these solutions were revisited in Ref.\cite{DuttaRoy:2019hij} where the authors showed the necessity of extra matter beyond phantom to support the geometries of these generalized solutions. In particular, the quasinormal modes were studied. Also, this metric was recently studied in the context of the braneworld, where the authors in \cite{Sharma:2021kqb} embedded these solutions in a five dimensional warp braneworld and this brings the possibility to the weak energy condition being satisfied. The stability of these generalized solutions considering axial gravitational perturbations was made in \cite{Roy:2021jjg}.

In this paper we study the possibility of non-relativistic quantum particles to traverse the generalized Ellis-Bronnikov wormholes by considering quantum effects, such as tunneling. We have used the generalized Ellis-Bronnikov wormhole metric and, as a result, the effective potential exhibits a barrier centered at the throat of the wormhole.

This paper is organized as follows: in the next section we present briefly the generalized Ellis-Bronnikov wormhole spacetime. In section III we found the Schr\"{o}dinger equation for a particle constrained to move along the generalized Ellis-Bronnikov wormhole spacetime and discuss the effective potential for different values of $n$ and orbital angular momentum $\ell$. In section IV we obtain analytical solutions for the Schr\"{o}dinger equation for $n=2$ and any value of orbital angular momentum. In section V, through a delta-barrier approximation we found the transmission and reflection coefficients for the $n=2$ case. In section VI we present our conclusions.

\section{Generalized Ellis-Bronnikov wormhole spacetime}

In this section we present a brief review of the Ellis-Bronikov wormhole and its generalized version.

\begin{figure*}[ht!]
    \centering
    \includegraphics[scale=0.8]{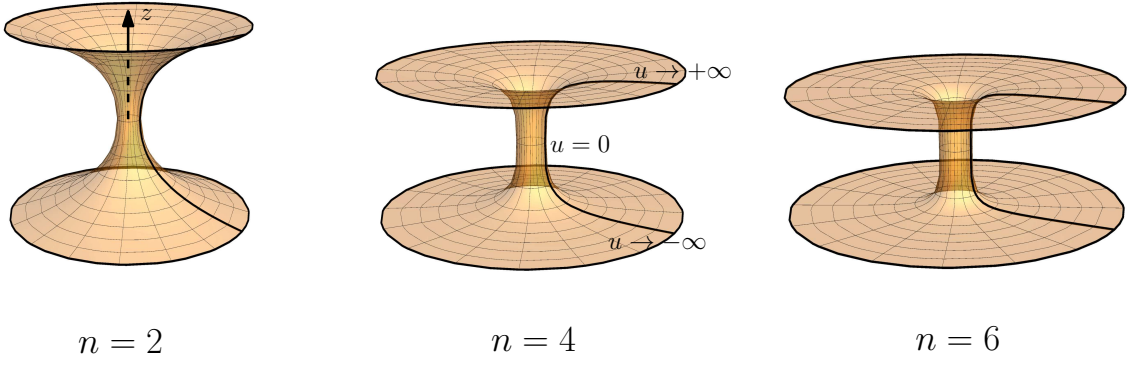}
    \caption{Wormhole's coordinate system and the effect of the $n$ deformation. The variable $u$ is highlighted in black.}
    \label{wormholefig}
\end{figure*}

Let us consider a static and spherically symmetric Morris-Thorne wormhole metric given by \cite{Morris:1988cz}
\begin{equation}\label{metric1}
    ds^2=e^{2\phi(r)}dt^2-\frac{dr^2}{1-\frac{b(r)}{r}}-r^2d\Omega_2,
\end{equation}
where $\phi(r)$ is the redshift function, $b(r)$ is the shape function and $d\Omega_2=d\theta^2+\sin\theta d\phi^2$ is the spherical line element. The Ellis-Bronnikov wormhole is defined by the following conditions on the shape and redshift functions \cite{Ellis:1973yv, Bronnikov:1973fh}
\begin{eqnarray}
b(r)&=&\frac{r_t^2}{r},\\
\phi(r)&=&0,
\end{eqnarray}
with $r_t$ being the wormhole's throat radius. The condition $\phi(r)=0$ imply in a zero tidal wormhole. Under these considerations the Morris-Thorne metric wormhole (\ref{metric1}) becomes
\begin{equation}\label{metric2}
    ds^2=dt^2-\frac{dr^2}{1-\frac{r_t^2}{r^2}}-r^2d\Omega_2.
\end{equation}
For our purpose it is more convenient to express the wormhole in a coordinate system in which we can distinguish between the upper and the lower plan. Hence, in terms of a new variable $u=\sqrt{r^2-r_t^2}$ (known as tortoise or proper radial distance coordinate) (highlighted in thick black in fig. (\ref{wormholefig})) we can rewrite the metric as
\begin{eqnarray}\label{EBreducedmetricu}
ds^2=dt^2-du^2-(u^2+r_t^2)d\Omega_2.
\end{eqnarray}

The generalized Ellis-Bronnikov wormhole spacetime is characterized by the following metric \cite{Kar:1995jz}
\begin{eqnarray}\label{GEBreducedmetric}
ds^2=dt^2-du^2-f^2(u)d\Omega_2,
\end{eqnarray}
with $f(u)$ given by
\begin{eqnarray}\label{f(u)}
f(u)=(u^n+r_t^n)^{1/n}.
\end{eqnarray}
The parameter $n$ is allowed to assume only even values, in order to guarantee the smooth behaviour of $f(u)$ over the entire domain. Note that $n=2$ recovers the usual Ellis-Bronnikov wormhole spacetime. Also, it is important to highlight that for $n=2$ we have a catenoid shaped wormhole while as we increase the value of $n$ the wormhole's shape approaches to a cylinder, as we can see in (\ref{wormholefig}).


\section{Non-relativistic Hamiltonian}

In this section we derive the non-relativistic Hamiltonian on a general wormhole spacetime and obtain the effective potential of a particle governed by the Schr\"{o}dinger equation.
Consider the generic metric
\begin{equation}
    ds^2 = -e^{2A(r)}dt^2 + \frac{1}{B(r)}dr^2 + r^2 (d\theta^2 + \sin^2 \theta d\phi^2).
\end{equation}
The Lorentz-invariant mass relation $g_{\mu\nu}P^{\mu}P^{\nu}=-(mc)^2$ in this spacetime yields to
\begin{equation}
    -e^{2A(r)}\left(\frac{E}{c}\right)^2 + g_{ij}P^{i}P^{j}=-(mc)^2.
\end{equation}
Then, the particle energy is related to the momentum $\Vec{p}$ by
\begin{equation}
    E=\pm e^{-A}(m^2c^4 + g_{ij}P^{i}P^{j})^{1/2}.
\end{equation}
In the Newtonian limit, $e^{-A}\approx 1-\frac{A}{c^2}+ \cdots$. Thus, by expanding the expression in powers of $P/mc$, we obtain
\begin{equation}
    E=\pm E_0 \pm \frac{P^2}{2m} \pm m A + \cdots,
\end{equation}
where $E_0 = mc^2$ and $P^2 =g_{ij}P^{i}P^{j}$. Defining the particle Hamiltonian as $H=E-E_0$ and considering the quantization $\hat{P}_i = -i\hbar \nabla_i$, where $\nabla_i$ is the covariant derivative, we obtain the non-relativistic Hamiltonian operator
\begin{equation}
    \hat{H}=-\frac{\hbar^2}{2m}g^{ij}\nabla_i \nabla_j + mA.
\end{equation}
It is worthwhile to mention that the wormhole geometry is present in the non-relativistic Hamiltonian by means of the space metric $g^{ij}$ and the redshift function $A$.

\begin{figure*}[ht!]
    \centering
\begin{subfigure}{.5\textwidth}
  \centering
  \includegraphics[scale=0.78]{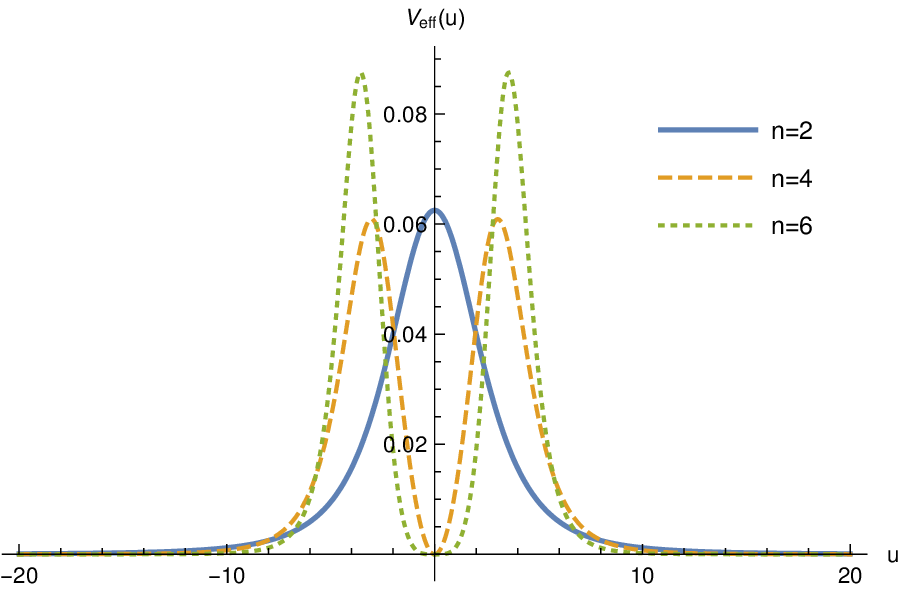}
  \caption{}

\end{subfigure}%
\begin{subfigure}{.5\textwidth}
  \centering
  \includegraphics[scale=0.78]{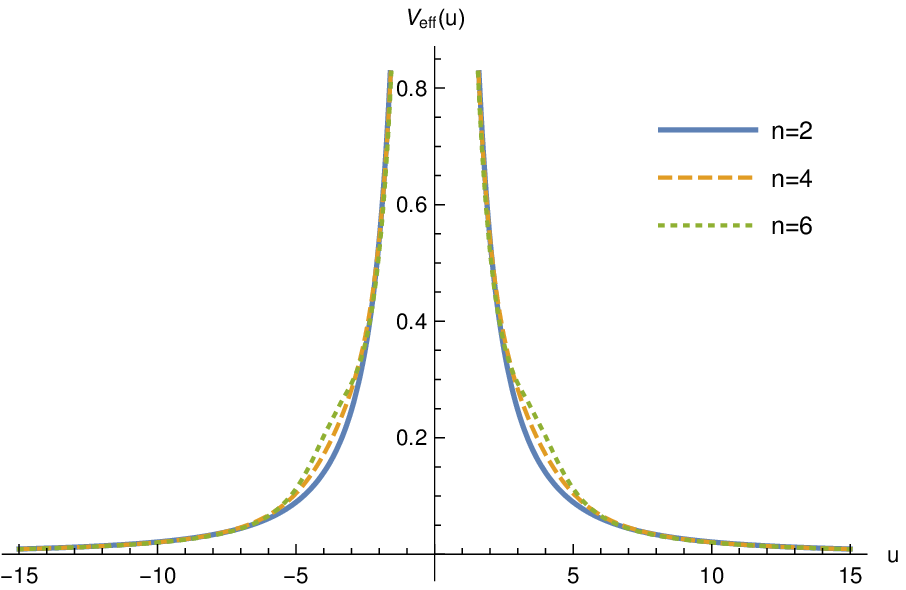}
  \caption{}

\end{subfigure}%
    \caption{Effective potential governing the dynamics of a non-relativistic particle in the generalized Ellis-Bronnikov wormhole spacetime. In (a) we have set $\ell=0$ while in (b) we have considered $\ell=1$. For both cases we set $r_t=4$.}
    \label{pot1}
\end{figure*}

Adopting the tortoise coordinate
\begin{eqnarray}
u=\int B^{-1/2}dr,
\end{eqnarray}
the metric can be written as $ds^2=-(1+\frac{A}{c^2})dt^2 + du^2 + f(u)^2(d\theta^2 +\sin{\theta}^2 d\phi^2)$, whereas the Hamiltonian operator takes the form
\begin{eqnarray}\label{schrodinger}
\hat{H}\Psi &=& -\frac{\hbar^2}{2m}\Bigg[\frac{\partial^2 \Psi}{\partial u^2} + \frac{1}{f^2(u)}\frac{\partial^2 \Psi}{\partial\theta^2}+\frac{1}{f^2(r)\sin^2{\theta}}\frac{\partial^2\Psi}{\partial\phi^2}\nonumber\\
& + & 2\frac{f'}{f}\frac{\partial\Psi}{\partial u}\Bigg] + m A(r)\Psi,
\end{eqnarray}
where $r=f(u)$ and the prime ' stands for $\frac{d}{du}$. Let us consider an expansion of the wave function on the spherical harmonics
\begin{equation}
    \Psi(u,\theta,\phi)=\sum \chi(u)Y^{m}_{l}(\theta, \phi),
\end{equation}
where the spherical harmonics satisfy $\nabla^2 Y^{m}_{l}(\theta,\phi)=-\frac{l(l+1)}{u^2}Y^{m}_{l}(\theta,\phi)$. Thus, the stationary Schr\"{o}dinger equation $\hat{H}\Psi = E\Psi$ leads to
\begin{eqnarray}
\chi'' + 2 \frac{f'}{f}\chi' + \left(mA(r) - \frac{l(l+1)}{u^2}\right)\chi=-\frac{2mE}{\hbar^2}\chi.
\end{eqnarray}

As highlighted in \cite{JEGSilva1,catenoid,Yesiltas}, the first-order derivative term breaks the hermiticity of the Hamiltonian operator. But since the Hamiltonian is invariant under $\mathcal{PT}$ symmetry, it is possible to find an Hermitian equivalent Hamiltonian with the same spectrum of eigenvalues which encodes all the geometrical information in an effective potential. Performing the change on the wave function $\Psi=\alpha \psi$, such that
$\alpha=f^{-1}$, we obtain
\begin{equation}
\label{radialequation}
    \psi''+\left(k^2 + mA -\frac{l(l+1)}{u^2} -\left(\frac{f'}{f}\right)' - \left(\frac{f'}{f}\right)^2 \right)\psi=0,
\end{equation}
where $k^2=\frac{2mE}{\hbar^2}$. Eq.(\ref{radialequation}) describes the radial propagation of the non-relativistic electron taking into account the redshift function $\phi$, the spherical symmetry and the deformed wormhole function $f(u)$.

From the Schr\"{o}dinger equation writen above we can write explicitly the effective potential governing the system as
\begin{equation}\label{effectivepotential}
    V_{eff}(u)=\frac{l(l+1)}{u^2} +\left(\frac{f'}{f}\right)' + \left(\frac{f'}{f}\right)^2-mA.
\end{equation}
Let us study now the effective potential for the generalized Ellis-Bronnikov wormhole spacetime, which is given by $A=0$ and $f(u)$ given by equation (\ref{f(u)}). In this case, the behaviour of the effective potential is depicted in fig.(\ref{pot1}).

This potential exhibits a barrier behaviour around the throat of the wormhole independent of the value of the orbital angular momentum $\ell$, as depicted in fig.(\ref{pot1}). For the orbital angular momentum $\ell=0$ fig.(\ref{pot1} a) the effective potential presents a single barrier centered at $u=0$ for $n=2$ and a symmetric double-barrier also centered at the $u=0$ for $n>2$. The presence of these symmetric double-barriers for $n>2$ is related to the fact that the increasing of $n$ promotes the emergence of two symmetric points of high curvature, where the throat connects with the flat spacetime regions. Also, as we increase the value of $n$ we increase the energy scale of the effective potential simultaneously to an increase in the distance between the barriers. For $\ell=1$ the effective potential exhibits a single barrier centered at $u=0$ for any value of $n$. In fact, for all values of $n$ the curves are practically overlapping, as we can see in fig.(\ref{pot1} b), except for a narrow region.

\section{Solutions}
\begin{figure}[h!]
    \centering
    \includegraphics[scale=0.3]{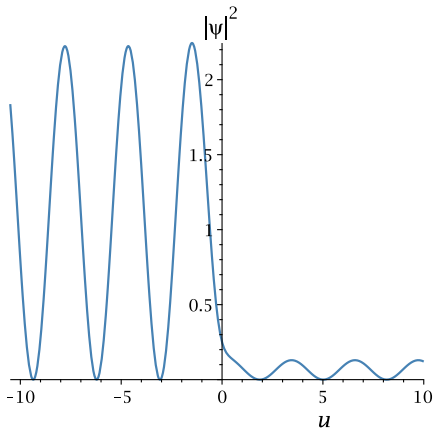}
   \caption{ Probability density of the wave function for a non-relativistic particle approaching the throat of an Ellis-Bronnikov $(n=2)$ wormhole for $\ell=0$, $k=1$, and $r_t=\meio$.}
    \label{fig3}
\end{figure}

There are no analytic solutions for Eq. \eqref{radialequation} unless $n=2$ (classical Ellis-Bronnikov wormhole spacetime). Then in this case we have (A=0)
\bea
\psi''(u)+\left(k^2-\frac{\ell (\ell+1)}{u^2} - \frac{r^2}{(r^2+u^2)^2}\right)\psi(u)=0.\label{eq19}
\eea
It is possible to show that the solutions for the above equation are the well-known confluent Heun functions ($H_C$) written as
\begin{eqnarray}
\nonumber \psi^{(1)}(u)&=&c_1 u^{-\ell}\sqrt{r_t^2+u^2}\times\\
&&\times H_C\left(0,-\ell-\meio,0, -\frac{k^2 r^2}{4}, \frac{k^2 r^2+1}{4}, -\frac{u^2}{r^2}\right)\nonumber\\
\end{eqnarray}
\begin{eqnarray}
\nonumber \psi^{(2)}(u)&=&c_2 u^{\ell+1}\sqrt{r_t^2+u^2}\times\\
&&\times H_C\left(0,\ell+\meio,0, -\frac{k^2 r^2}{4}, \frac{k^2 r^2+1}{4}, -\frac{u^2}{r^2}\right)\nonumber\\
\end{eqnarray}
where $c_1$ and $c_2$ are normalization constants. Since the effective potential vanishes in the asymptotic limits, wave function solutions for $V_{eff}\rightarrow 0$ yields a plane wave solution, therefore describing a free particle. The free particle description in the asymptotic limits is in agreement with the fact that the wormhole is asymptotically flat. As the particle approaches the throat, the effects of the curvature becomes relevant as shown by the confluent Heun functions. The probability density of the wave function as the particle approaches the throat of the wormhole is presented in figure (\ref{fig3}).

As we can see, the figure (\ref{fig3}) exhibits an scattering through the barrier located at $u=0$. Considering a particle coming from $u\rightarrow -\infty$, the figure (\ref{fig3}) suggests a small rate of transmission through the barrier.

\section{Delta-barrier approximation}

For $n=2$ and $l=0$, the potential given by Eq. (\ref{effectivepotential}), considering a small throat radius, $r_t\to 0$, and at a region nearby the throat $u\approx r_t$ can be approximated as being proportional to the delta function, $V_{eff} \approx \lambda \delta(u)$. The constant $\lambda$ can be computed via
\begin{equation}
\lambda\int_{-\infty}^{\infty} \delta(u)du\approx \int^{\infty}_{-\infty}\frac{r_t^2}{(u^2+r_t^2)^2} du\Rightarrow \lambda\approx \frac{\pi}{2 r_t}.
\end{equation}

Thus, the Schr\"{o}dinger equation can be easily solved, allowing one calculates the transmission ($T$) and reflection ($R$) coefficients of quantum tunnelling of a particle with mass $m$ and energy $E<V_{eff}^{max}=1/r_t^2 $ through the wormhole throat, which are given by
\begin{eqnarray}
    T&\approx&\frac{1}{1+\frac{mG\hbar\pi^2}{8 c r_t^2E}},\\
    R&\approx&\frac{1}{1+\frac{8cr_t^2E}{m G\hbar\pi^2}},
\end{eqnarray}
where we have restored the fundamental constants. Notice that $R+T\approx1$.

\section{Final Remarks}

In this paper we study the possibility of non-relativistic quantum particles to traverse the generalized Ellis-Bronnikov wormholes by considering quantum effects, such as tunneling. 

We have used the generalized Ellis-Bronnikov wormhole metric and found that for $n=2$ we have a single barrier shaped effective potential centered at the throat of the wormhole for any value of orbital angular momentum. For $n\neq2$ we have a symmetric double barrier shaped potential also for orbital angular momentum $\ell=0$. As we increase the value of $n$ we also increase the separation distance between the barriers when $\ell=0$. For $\ell\neq 0$ we have always a barrier centered at $u=0$ for any value of $n$.  

Analytical solutions for the Schr\"{o}dinger equation in the generalized Ellis-Bronnikov spacetime could be found only for $n=2$. Such solutions were given in terms of the confluent Heun functions. 

Finally, by using a delta-barrier approximation we could find the transmission and reflection coefficients for a non-relativistic particle to traverse the generalized Ellis-Bronnikov wormhole in terms of the wormhole parameters and the energy of the incident particle.


\begin{thebibliography}{99}

\bibitem{Einstein:1935tc}
A.~Einstein and N.~Rosen,
Phys. Rev. \textbf{48}, 73-77 (1935)
doi:10.1103/PhysRev.48.73

\bibitem{Morris:1988cz}
M.~S.~Morris and K.~S.~Thorne,
Am. J. Phys. \textbf{56}, 395-412 (1988)
doi:10.1119/1.15620

\bibitem{Sushkov:2005kj}
S.~V.~Sushkov,
Phys. Rev. D \textbf{71}, 043520 (2005)
doi:10.1103/PhysRevD.71.043520
[arXiv:gr-qc/0502084 [gr-qc]].

\bibitem{Lobo:2005us}
F.~S.~N.~Lobo,
Phys. Rev. D \textbf{71}, 084011 (2005).

\bibitem{Garattini:2019ivd}
R.~Garattini,
Eur. Phys. J. C \textbf{79}, no.11, 951 (2019)

\bibitem{Jusufi:2020rpw}
K.~Jusufi, P.~Channuie and M.~Jamil,
Eur. Phys. J. C \textbf{80}, no.2, 127 (2020).

\bibitem{Alencar:2021ejd}
G.~Alencar, V.~B.~Bezerra and C.~R.~Muniz, Eur. Phys. J. C {\bf 81}, 924 (2021).

\bibitem{Oliveira:2021ypz}
P.~H.~F.~Oliveira, G.~Alencar, I.~C.~Jardim and R.~R.~Landim,

\bibitem{Carvalho:2021ajy}
I.~D.~D.~Carvalho, G.~Alencar and C.~R.~Muniz, Int. J. Mod. Phys. {\bf D} 31, No. 03, 2250011 (2022).

\bibitem{Richarte:2007zz}
M.~G.~Richarte and C.~Simeone,
Phys. Rev. D \textbf{76}, 087502 (2007).
[erratum: Phys. Rev. D \textbf{77}, 089903 (2008)].

\bibitem{Matulich:2011ct}
J.~Matulich and R.~Troncoso,
JHEP \textbf{10}, 118 (2011).

\bibitem{Richarte:2009zz}
M.~G.~Richarte and C.~Simeone,
Phys. Rev. D \textbf{80}, 104033 (2009)
[erratum: Phys. Rev. D \textbf{81}, 109903 (2010)].

\bibitem{MontelongoGarcia:2011ag}
N.~Montelongo Garcia and F.~S.~N.~Lobo,
Mod. Phys. Lett. A \textbf{40}, 3067-3076 (2011)

\bibitem{Ovgun:2018xys}
A.~\"Ovg\"un, K.~Jusufi and \.I.~Sakall\i{},
Phys. Rev. D \textbf{99}, no.2, 024042 (2019).

\bibitem{Lessa}
L.~A.~Lessa, R.~Oliveira, J.~E.~G.~Silva and C.~A.~S.~Almeida,
Annals Phys. \textbf{433} (2021), 168604.

\bibitem{Chew:2016epf}
X.~Y.~Chew, B.~Kleihaus and J.~Kunz,
Phys. Rev. D \textbf{94}, no.10, 104031 (2016).

\bibitem{Chew:2018vjp}
X.~Y.~Chew, B.~Kleihaus and J.~Kunz,
Phys. Rev. D \textbf{97}, no.6, 064026 (2018).

\bibitem{Ellis:1973yv}
H.~G.~Ellis,
J. Math. Phys. \textbf{14}, 104-118 (1973).

\bibitem{Bronnikov:1973fh}
K.~A.~Bronnikov,
Acta Phys. Polon. B \textbf{4}, 251-266 (1973)

\bibitem{Kar:1995jz}
S.~Kar, S.~Minwalla, D.~Mishra and D.~Sahdev,
Phys. Rev. D \textbf{51}, 1632-1638 (1995)

\bibitem{DuttaRoy:2019hij}
P.~Dutta Roy, S.~Aneesh and S.~Kar,
Eur. Phys. J. C \textbf{80}, no.9, 850 (2020)

\bibitem{Sharma:2021kqb}
V.~Sharma and S.~Ghosh,
Eur. Phys. J. C \textbf{81}, no.11, 1004 (2021).

\bibitem{Roy:2021jjg}
P.~D.~Roy,
[arXiv:2110.05019 [gr-qc]].

\bibitem{JEGSilva1}
J.E.G. Silva et al., Phys. Lett. A {\bf 384}, 126458 (2020).

\bibitem{catenoid}
J.E.G. Silva, J. Furtado and A.C.A. Ramos, 
Eur. Phys. J. B 94, 127 (2021).

\bibitem{Yesiltas}
\"O.~Ye\c{s}ilta\c{s}, J.~Furtado and J.~E.~G.~Silva,
Eur. Phys. J. Plus \textbf{137} (2022) no.4, 416.

\end{thebibliography}
\end{document}